# ENERGY SPREAD OF THE PROTON BEAM IN THE FERMILAB BOOSTER AT ITS INJECTION ENERGY *

C. M. Bhat[#], B. E. Chase, S. J. Chaurize, F. G. Garcia, K. Seiya,
W. A. Pellico, T. M. Sullivan, A. K. Triplett, Fermilab, Batavia, IL 60510, USA

## Abstract

We have measured the total energy spread (≈99% energy spread) of the Booster beam at its injection energy of 400 MeV by three different methods - 1) creating a notch of about 40 nsec wide in the beam immediately after multiple turn injection and measuring the slippage time required for high and low momentum particles for a grazing touch in line-charge distribution, 2) injecting partial turn beam and letting it to debunch, and 3) comparing the beam profile monitor data with predictions from MAD simulations for the 400 MeV injection beam line. The measurements are repeated under varieties of conditions of RF systems in the ring and in the beam transfer line.

## INTRODUCTION

The Booster is the oldest proton synchrotron in the Fermilab accelerator complex, operating at 8 GeV extraction energy since May 1971 [1, 2] (it was tested for 10 GeV extraction energy for a short time). Though it is a rapid cycling synchrotron (sinusoidal magnet ramp at 15 Hz with a frequency variation of ≈ 0.2% variation) the beam is not delivered on all of its cycles, all the time. Many improvements have been made over the past four decades to increase the beam intensity per Booster *batch* and to increase its average beam power to the down-stream accelerators and to the low energy neutrino program — full potential of the Booster is yet to be realized. One of the important goals of the "Proton Improvement Plan" at Fermilab [3] is to extract the beam at 15 Hz rate from the Booster all the time. Long range plan of Fermilab [4] is to increase the Booster beam delivery cycle rate from 15 Hz to 20 Hz. Hence, the current Booster plays a very significant role in the near future of the Fermilab.

In support of the proposed upgrades to the Fermilab facilities, a thorough investigation on the properties of the beam at the injection energy is extremely valuable. In the past, many attempts have been made in this regard (e.g., ref. [5]), in bits and pieces, often only soon after major upgrades in the upstream hardware. Nevertheless, we realized that a systematic and coherent set of measurements on beam properties at injection is highly needed to understand the beam dynamics.

The Booster receives 400 MeV (kinetic energy) proton beam from the LINAC with 200 MHz bunch structure, while it is at its minimum of sinusoidal ramp. The injection line is equipped with an 800 MHz debuncher cavity to reduce the energy spread of the injected beam. After the injection the beam is quickly captured with the help of a 37 MHz RF system. Typically the capture efficiency is <97 %. Between the injection and RF capture the beam is debunched for ~100 μs. In this report, we primarily focus on the measurement of energy spread at this time of the beam cycle as a function of beam intensity which is measured in terms of Booster Turns (BT) (1BT≈0.33E12 protons).

## SIMULATIONS AND MEASUREMENTS

In the absence of any RF field, injected bunched beam into the synchrotron debunches (shears) forming a DC beam and maintains its initial energy spread. The energy spread of such a beam can be measured by producing a gap (notch) in the beam with the help of a fast kicker and measuring the time $T_{Graz}$ required for a grazing touch of two sides of its line-charge distribution. Then the energy spread $\Delta E$ is given by,

$$\Delta E = \frac{\beta^2 E_s}{|\eta|} \frac{W_{notch}}{T_{Graz}} \qquad (1)$$

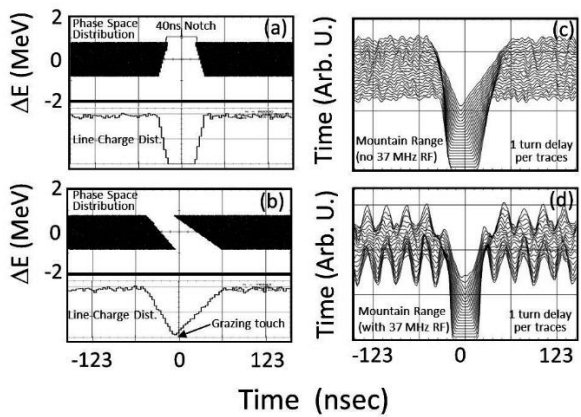

Figure 1: ESME[6] simulations for a) initial beam phase space distribution with a notch in the beam (top) and its line-charge distribution (bottom) b) same as "a" but at grazing touch, c) mountain range without 37 MHz RF and c) similar to "c" but with 37 MHz RF turned on at 25kV.

where $\beta$, $E_s$, $\eta$ and $W_{notch}$ are relativistic velocity (0.713 for 400 MeV proton), synchronous energy of the beam (1338.3 MeV), slip factor (-0.4583) and the width of the notch in seconds (≈40 nsec), respectively. We have performed longitudinal beam dynamics simulations for the Booster injection scenario using the machine parameters [2] and for a notch of ≈40 ns width to understand the notch filling mechanism. Fig. 1 depicts the predicted dynamics of the beam particle distribution and the mountain range for this process. The shape of the notch as an input to the simulation is derived from scope data. For a $\Delta E$=1.6 MeV, it takes about 21 revolutions for the grazing touch. In the presence of a small amount of RF the dynamics of notch filling is more complicated as shown in Fig. 1(d). Nevertheless, it takes the same number of revolutions to fill the gap.

___________________________________________
* Work supported by Fermi Research Alliance, LLC under Contract No. De-AC02-07CH11359 with the United States Department of Energy
# cbhat@fnal.gov

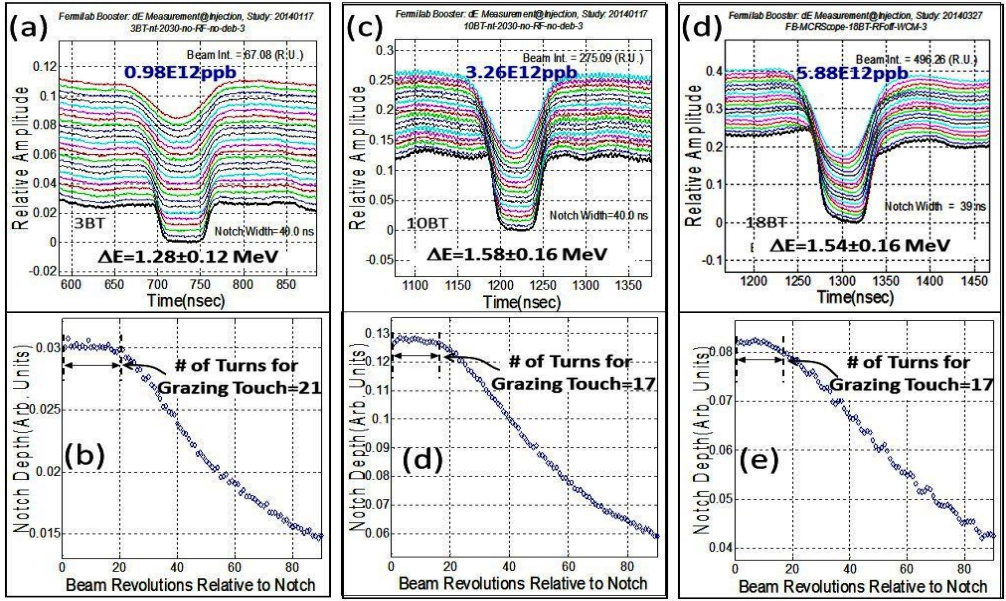

Figure 2: Measured mountain range data and relative notch depth as a function of number of revolutions beam makes as it is being filled. The notch widths in all these cases were ~40 ns (95%). (a) and (b) are for 0.98E12ppb,(c) and (d) are for 3.26E12ppb and (e) and (f) are for 5.88E12ppb. These data are with Booster and debuncher RF turned off.

The experiments have been conducted under varieties of conditions: 1) different beam intensities, i.e., (i) partial BT, (ii) 2-18 BT, 2) Booster RF turned off, 3) Booster RF turned on but vector sum of RF voltage set to zero, 4) the debuncher RF cavity turned off and 5) the debuncher RF cavity turned on. After the beam injection a notch of about 40 nsec (FWHM) is produced in the beam using a set of fast kickers except in the case of 1(i). The evolution of line-charge distribution on every turn is measured for the first 400 µs after the beam injection using beam signals from a wall current monitor (WCM) fed to a Tektronix TDS5054B-NV-T 500MHz, 5GSa/s, 4ch Digital Oscilloscope and triggering the scope on the beam event. The time required to fill up the notch as explained in the simulations or beam slippage rate is extracted from the mountain range data.

## RESULTS AND DISCUSSIONS

A MATLAB software program has been developed for off-line scope data analysis which extracts the energy spread using constructed mountain range plots. The WCM monitor used for these measurements is an AC coupled system. Hence, if the beam is fully debunched and does not have any notch (or RF) then the line-charge distribution is a uniform distribution. The moment a notch is produced a clear indentation can be seen depending on the extent of the kick. Figures 2(a), (c) and (d) show measured mountain range for three different beam intensities with Booster and debuncher RF cavity turned off. The exact shape of a notch soon after it is formed in each case is shown by the lowest black trace. Figures 2(b), (d) and (e) are the measured depths of notches as they get filled up due to slippage of the high and low momentum protons. $T_{Graz}$ is a measure of number of revolutions needed from the start of the notch to the grazing touch as indicated in each figure.

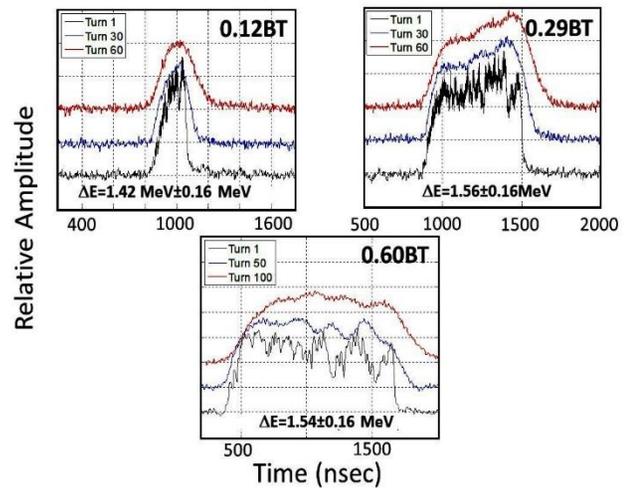

Figure 3: Measured line-charge distributions for partial turn beam with Booster and debuncher RF cavities RF off.

Figure 3 shows the WCM data for the partial beam in the Booster as they debunch. For clarity we have lined up the rising edges of the WCM data to measure the slippage of the beam particles due to shearing.

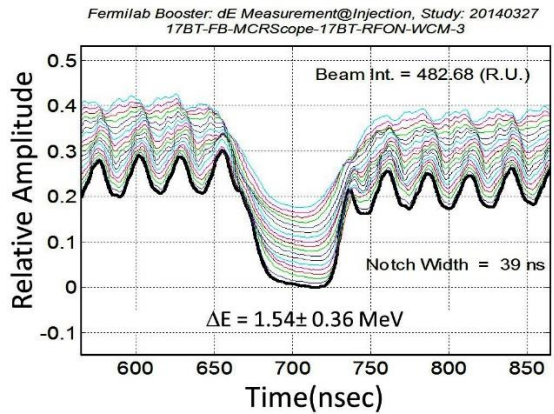

Figure 4: Notch in a 17BT beam with the Booster RF on at ~30kV.

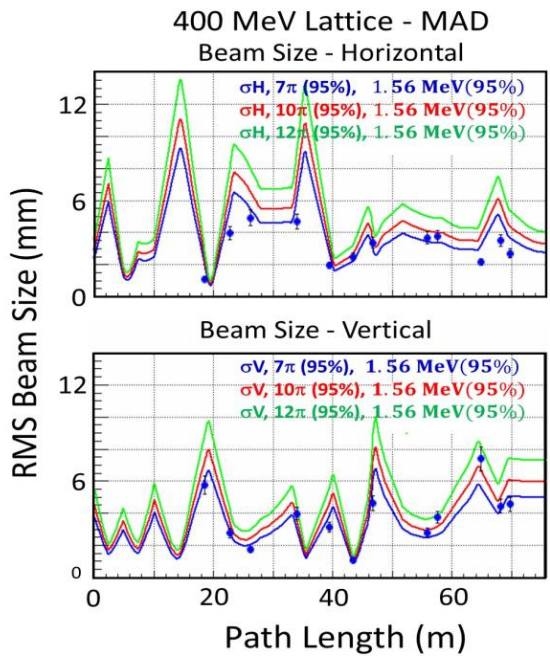

Figure 5: Measured beam size using multi-wire and its comparison with MAD predictions for the 400 MeV injection beam line.

We expect about 10% error in the measured energy spread for the cases with no RF in the Booster. This error mainly arises in estimating number of revolution needed for grazing touch. Figure 4 shows a sample of mountain range data with 37 MHz RF of ≈ 30 kV. The error in this case is rather large as compared with no RF.

Table 1 summarizes our results for various scenarios. We conclude that the average energy spread in the Fermilab Booster at injection energy is less than 1.50±0.16 MeV and its dependence on booster turns is quite weak; the central value found to vary in the range 1.28 to 1.80 for 0.1 BT to 18BT. Also, during these measurements the debuncher appeared to have little effect on the beam energy spread, which indicates that we still need better optimization of the debuncher cavity parameters to take advantage of it in reducing the energy spread in injected beam.

We have also estimated the beam energy spread in the 400 MeV beam by measuring the beam sizes using 12 horizontal and vertical multi-wire beam profile monitors and comparing the data with MAD predictions of the beam line lattice function. (Notice that the effect of the debuncher cavity is included in the analysis). The results are shown in Fig. 5. The optimum energy spread obtained by this method is consistent with that obtained by creating a notch in the Booster beam.

Furthermore, we plan to develop a console application program which uses the method adopted to measure the beam energy spread at injection by creating a notch. This might immensely help the future operations.

Table 1: Measured beam energy spread in MeV for partial and full Booster turns.

| A*     | B*     | 3BT   | 10BT  | 13BT  | 17BT  | 18BT  |
|--------|--------|-------|-------|-------|-------|-------|
| RF ON  | RF ON  | 1.42 ±0.30 | 1.32 ±0.30 | 1.46 ±0.32 | 1.46 ±0.32 | 1.48 ±0.32 |
| RF OFF | RF ON  | 1.38 ±0.12 | 1.40 ±0.14 | 1.46 ±0.16 | 1.40 ±0.14 | 1.48 ±0.16 |
| RF OFF | RF OFF | 1.38 ±0.12 | 1.5 ±0.14 |       | 1.5 ±0.14 |       |

For Partial Turn Beam

| BT   | A*     | B*     | (MeV)     |
|------|--------|--------|-----------|
| 0.6  | RF ON  | RF ON  | 1.72 ±0.38 |
|      | RF OFF | RF ON  | 1.54 ±0.12 |
|      | RF OFF | RF OFF | 1.84 ±0.12 |
| 0.35 | RF OFF | RF OFF | 1.40 ±0.2 |
| 0.29 | RF OFF | RF OFF | 1.56 ±0.2 |
| 0.12 | RF OFF | RF OFF | 1.42 ±0.2 |

**A** stands for Booster RF cavity and **B** is for Debuncher RF cavity in the 400 MeV beam line.

Authors would like to thank the MCR crew for their help during these beam measurements.

## REFERENCES

[1] R. R. Wilson, Proc. of the 8th Int. Conf. on High Energy Accel. 1971, Geneva, Switzerland, p3; R. Billinge et. al, IEEE Trans. Nuc. Sci. 1969 Particle Accel. Conf. p. 969.
[2] Booster Rookie Book, http://www-bdnew.fnal.gov/ operations/rookie_books/ rbooks.html.
[3] W. Pellico, et al., IPAC2014 (2014) p 3409; R. Webber et al., Beams Document 3781-v2 (2011).
[4] "The PIP-II Reference Design Report," V1 March 2015, http://www-bdnew.fnal.gov/pxie/PIP-II_RDR/ PIP-II_RDR.pdf.
[5] M. Popovic, (private communications, 2014).
[6] J. MacLachlan, http://www-ap.fnal.gov/ESME/.